\pdfoutput=1
\documentclass[11pt,a4paper]{article}

\usepackage[margin=25mm]{geometry}
\usepackage{amsmath,amssymb,amsthm,mathtools,amscd}
\usepackage{array,booktabs,enumitem}
\usepackage[
  colorlinks=true,
  linkcolor=blue,
  citecolor=blue,
  urlcolor=blue
]{hyperref}

\newtheorem{definition}{Definition}[section]
\newtheorem{proposition}[definition]{Proposition}
\newtheorem{remark}[definition]{Remark}

\newcommand{\Econ}{\mathsf{Econ}}
\newcommand{\Prob}{\mathsf{Prob}}
\newcommand{\Cat}{\mathsf{Cat}}
\newcommand{\Graph}{\mathsf{Graph}}
\newcommand{\Hist}{\mathsf{Hist}}
\newcommand{\FreeCat}{\operatorname{FreeCat}}
\newcommand{\Tw}{\operatorname{Tw}}
\newcommand{\Hol}{\operatorname{Hol}}
\newcommand{\id}{\operatorname{id}}
\newcommand{\Tmic}{\mathcal T_{\mathrm{mic}}}
\newcommand{\Tmac}{\mathcal T_{\mathrm{mac}}}
\newcommand{\Hmic}{\mathcal H_{\mathrm{mic}}}
\DeclareMathOperator{\src}{src}
\DeclareMathOperator{\tgt}{tgt}

\title{Toward an Admissibility Theory of Aharonov--Bohm Arbitrage\\
\large A working note from elementary examples}
\author{Takanori Adachi\footnote{Graduate School of Management, Tokyo Metropolitan University, Marunouchi Eiraku Bldg. 18F, 1-4-4 Marunouchi, Chiyoda-ku, Tokyo 100-0005 Japan\\Email: taka.adachi@tmu.ac.jp},
Keisuke Hara\footnote{Association of Mathematical Finance Laboratory},
\footnote{This work was supported by JSPS KAKENHI Grant Number 24K04941.}\footnote{
MSC-class:
    91-10, 
    91G15, 
    91G80, 
    81Q70 
}
}

\date{\today}

\begin{document}
\maketitle

\begin{abstract}
\noindent
This note develops a provisional admissibility framework for
Aharonov-Bohm type arbitrage. Starting from a categorical filtration
$
  F:\mathcal T_{\mathrm{mic}}^{\mathrm{op}}\to \mathsf{Prob}
$
and its distortion on microscopic histories, we ask when a favorable
closed microscopic loop can be realized as an executable relative
ordinal arbitrage over an open macro-time interval. The history
endofunctor \(\mathsf{Hist}\) retains the staged decomposition of economic
activities, while the twisted-arrow category
\(\operatorname{Tw}(\mathsf{Hist}(\mathcal T_{\mathrm{mic}}))\) represents a microscopic loop as an arrow between
distinct histories. An observable clock
\(\Pi: \mathsf{Econ} \to \mathcal T_{\mathrm{mac}} \) and a microscopic realization functor
$
  J: \mathsf{Hist}(\mathcal T_{\mathrm{mac}}) \to \operatorname{Tw}(\mathsf{Hist}(\mathcal T_{\mathrm{mic}}))
$
connect economic histories with their microscopic geometry.
Executability is treated separately through configuration histories,
capacity constraints, observations, and policies. This yields pathwise
and almost-sure notions of execution.
A finite foreign-exchange model illustrates stochastic execution, capacity failure, terminal settlement,
and comparison with a benchmark strategy. 
A parameterized finite filtration is constructed whose nontrivial
statewise distortion agrees exactly with the relative FX settlement
multiplier. Hence the resulting filtration--policy model satisfies the bridge
$V_*(\pi^q)=dF(\gamma)V_*(\pi^p),$
and, under order compatibility, realizes relative ordinal arbitrage.
The remaining problem is to identify the filtration parameters from
independent market primitives and to study stability under refinement.
\end{abstract}

\section{Introduction}
\label{sec:introduction}

The Aharonov--Bohm approach to arbitrage associates a multiplicative
distortion with a closed history in a categorical filtration
\cite{AdachiHaraAB}.  A favorable distortion detects a geometric failure of
path independence: transport around a microscopic loop returns to the same
time object but need not return the unit random variable.  This gives a
notion of geometric AB-arbitrage that is intrinsic to the filtration.

Such a loop is not yet a classical trading opportunity.  A classical
arbitrage is carried out between distinguishable initial and terminal times.
Its actions must be selected without using unavailable future information,
must remain feasible under resource and market-capacity constraints, and
must be settled after all economically relevant liabilities have been
included.  Moreover, the favorable microscopic distortion must explain the
economic advantage of the realized strategy.  These requirements constitute
the \emph{admissibility problem} considered in this note.

The difficulty has both a temporal and an economic aspect.  Temporally, a
functor sends a loop to a loop and therefore cannot directly turn a closed
microscopic history into an open macro-time interval.  Economically, the
existence of a formal activity does not imply that it can be executed from a
given configuration, and a favorable random multiplier does not by itself
identify a strategy that earns it.  The purpose of the present construction
is to separate these issues and state precisely where an additional bridge
is required.

We use two time categories.  The category \(\Tmic\) carries the microscopic
categorical filtration, whereas \(\Tmac\) records the observable dates and
intervals of economic activity.  The economic layer is described by a
preorder-enriched category \(\Econ\), together with an observable clock
\(\Pi:\Econ\to\Tmac\).  The formal definition of this structure, including
its liability-aware comparison of parallel activities, is given in
Section~\ref{sec:economy-category}.

Several constructions are introduced to connect these layers.  First, the
history endofunctor \(\Hist\) retains a composable string of activities before
its synthesis into a single categorical composite.  This distinction is
essential because executability, information availability, and distortion
may depend on the staged history.  Second, the twisted-arrow category
\(\Tw(\Hist(\Tmic))\) opens a microscopic loop into an arrow whose source and
target are distinct histories.  A realization functor
\[
  J:\Hist(\Tmac)\longrightarrow\Tw(\Hist(\Tmic))
\]
can therefore associate an open macro-time history with a closed microscopic
loop without requiring a direct projection from microscopic time to macro
time.  If \(\tau\) is a macro history, its realized distortion is the derived
quantity
\[
  dF_J(\tau)
  :=dF\bigl(\tgt(J(\tau))\bigr).
\]

Third, formal economic histories are separated from their executions.  An
execution projection records which configuration histories implement a
given economic history.  Observations and policies then specify actions as
information becomes available.  This gives pathwise and almost-sure notions
of executability and permits explicit capacity failures.  The information
used by a policy is allowed to be subjective and need not increase with
time; previously observed information may later be forgotten or represented
in a distorted form.

These constructions lead to five provisional admissibility conditions:
observable temporal compatibility, geometric realizability, predictable
selection, executable and complete economic realization, and compatibility
between microscopic distortion and economic comparison.  The first four
can be formulated using the categorical and execution data developed below.
The fifth is the central unresolved issue.  For a benchmark policy \(\pi^p\)
and a candidate policy \(\pi^q\), one seeks an independently justified
settlement bridge of the form
\[
  V_*(\pi^q)=dF(\gamma)V_*(\pi^p),
\]
together with compatibility between settlement-value dominance and the
hom-preorder of \(\Econ\).  Under such a bridge, a geometric AB-arbitrage
produces a relative ordinal arbitrage
\[
  e(\pi^p)\prec e(\pi^q).
\]

A finite foreign-exchange model is used as an axiom test.  It makes the
macro schedule, capacity constraints, observations, stochastic execution
lifts, and terminal settlement explicit, and it exhibits a genuine
capacity-failure event.  A parameterized non-flat filtration, modeled on
Example~7.2 of \cite{AdachiHaraAB}, realizes the two statewise FX settlement
ratios as the holonomy of a three-leg microscopic loop.  This completes the
settlement bridge at the level of finite representability.  The stronger
problem of deriving the filtration parameters from independent market
primitives remains open.

Accordingly, the aim of this working note is not to present a completed
characterization of admissible AB-arbitrage.  Its contribution is to replace
an informal admissibility requirement by a sequence of typed, separately
testable questions.  In schematic form, the intended implication is
\[
  \text{microscopic AB-loop}
  \quad\Longrightarrow\quad
  \text{executable relative ordinal arbitrage over macro-time},
\]
and the note gives a finite model of this implication while isolating the
additional naturality and refinement questions that remain open.

The organization is as follows.
Section~\ref{sec:economy-category} defines the economy category, its
hom-preorders, relative ordinal arbitrage, and the observable clock.
Section~\ref{sec:hidtoryEndofunctor}
introduces history categories
and the synthesis transformation.  
Sections~\ref{sec:microscopic-distortion},
\ref{sec:globalUnfolding}
and  \ref{sec:macro-micro-realization}
define microscopic distortion, the global twisted-arrow unfolding, 
and the micro--macro realization.  
Section~\ref{sec:admissibility}
formulates provisional admissibility conditions and tests them against
elementary counterexamples.
Section~\ref{sec:executability}
develops execution structures, observations, policies, and stochastic executability.  
Section~\ref{sec:fxExample}
studies the finite foreign-exchange model and constructs its settlement bridge.  
Section~\ref{conclusion}
summarizes the resulting framework and its remaining open problems.

\section{The economy category and observable macro time}
\label{sec:economy-category}

We now specify the economic layer used throughout the note.  Recall that a
category enriched over \(\mathsf{Preord}\) is an ordinary category
\(\mathcal C\) equipped with a preorder on every hom-set
\(\mathcal C(x,y)\), such that composition is monotone in both variables:
whenever
\[
  f\preceq f'\quad\text{and}\quad g\preceq g',
\]
for composable parallel arrows, one has
\[
  g\circ f\preceq g'\circ f'.
\]

\begin{definition}[Economy category]
An \emph{economy category} \(\Econ\) is a category enriched over
\(\mathsf{Preord}\) with the following interpretation.
\begin{enumerate}[label=\textnormal{(\roman*)},leftmargin=10mm]
\item An object \(x\) is a dated economic interface specifying the boundary
conditions under which economic activities may begin, end, and compose.
\item A morphism \(a:x\to y\) is a formally specified economic activity from
\(x\) to \(y\).  Its economic assessment includes every receipt, payment,
transaction cost, collateral requirement, and contractual obligation that
can affect its comparison with another activity.
\item For \(a,b\in\Econ(x,y)\), the relation
\[
  a\preceq b
\]
means that \(b\) is economically no worse than \(a\) under the common
boundary interfaces \(x\) and \(y\), after all relevant benefits and
liabilities have been taken into account.
\end{enumerate}
\end{definition}

Write \(a\sim b\) when \(a\preceq b\) and \(b\preceq a\), and define the
strict part of the preorder by
\[
  a\prec b
  \quad\Longleftrightarrow\quad
  a\preceq b\ \text{ and }\ b\not\preceq a.
\]
If \(a\prec b\), we call \(b\) a \emph{relative ordinal arbitrage over}
\(a\).  The benchmark \(a\) need not be a zero or null strategy, and no
addition or subtraction is imposed on the hom-sets of \(\Econ\).

The enriched category records formal economic comparison.  It does not by
itself assert that an activity can be implemented from a particular
configuration, or that a composable string remains feasible under capacity
constraints.  These execution questions are introduced separately in
Section~\ref{sec:executability}.

\begin{definition}[Observable clock]
The observable temporal placement of economic interfaces and activities is
specified by a functor
\[
  \Pi:\Econ\longrightarrow\Tmac.
  \tag{P}
\]
Thus an interface \(x\) has observable macro date \(\Pi(x)\), and an activity
\(a:x\to y\) has macro-time arrow
\[
  \Pi(a):\Pi(x)\longrightarrow\Pi(y).
\]
\end{definition}

The clock \(\Pi\) belongs to the observable market structure available to
economic agents.  Its use does not require an agent to recognize the
microscopic loop or the realization functor introduced later.

\section{The history endofunctor and synthesis}
\label{sec:hidtoryEndofunctor}

Let
\[
  U:\Cat\longrightarrow\Graph
\]
be the underlying directed-graph functor which forgets composition,
and let
\[
  \FreeCat:\Graph\longrightarrow\Cat
\]
be the free-category functor in which compositions are defined as concatenations. 
Define
\[
  \Hist:=\FreeCat\circ U:\Cat\longrightarrow\Cat.
\]
Thus, for a small category \(\mathcal C\),
\[
  \Hist(\mathcal C)=\FreeCat\bigl(U(\mathcal C)\bigr).
\]
The objects of \(\Hist(\mathcal C)\) are those of \(\mathcal C\). A
morphism \(\alpha:x\to y\) is a finite composable history
\[
  \alpha=
  \bigl(
  x=x_0\xrightarrow{f_1}x_1
  \xrightarrow{f_2}\cdots
  \xrightarrow{f_n}x_n=y
  \bigr).
\]
Composition in \(\Hist(\mathcal C)\) is concatenation of histories, and
the empty history \(\varepsilon_x\) is the identity at \(x\).

For a functor \(G:\mathcal C\to\mathcal D\), define
\[
  \Hist(G):
  \Hist(\mathcal C)\longrightarrow\Hist(\mathcal D)
\]
componentwise:
\[
  \Hist(G)
  \bigl(
  	x_0
		\xrightarrow{f_1}
	\cdots
  		\xrightarrow{f_n}
	x_n
  \bigr)
=
  \bigl(
  	G(x_0)
		\xrightarrow{G(f_1)}
	\cdots
  		\xrightarrow{G(f_n)}
	G(x_n)
  \bigr) .
\]
This construction preserves identities and composition of functors, so
\(\Hist\) is an endofunctor on \(\Cat\).

For each \(\mathcal C\), categorical composition gives a synthesis functor
\[
  Q_{\mathcal C}:
  \Hist(\mathcal C)\longrightarrow\mathcal C,
  \qquad
  Q_{\mathcal C}(f_1,\ldots,f_n)
  =
  f_n\circ\cdots\circ f_1.
\]
These functors are natural in \(\mathcal C\). Namely, for every
\(G:\mathcal C\to\mathcal D\),
\[
  Q_{\mathcal D}\circ\Hist(G)
  =
  G\circ Q_{\mathcal C}.
  \tag{N}
\]
Therefore,
\[
  Q:\Hist\Longrightarrow\operatorname{Id}_{\Cat}
\]
is a natural transformation. In fact, since
\(\FreeCat\dashv U\), the endofunctor \(\Hist=\FreeCat\circ U\)
carries a comonad structure and \(Q\) is its counit.

For the economy category, a history
\[
  p:x\longrightarrow y
  \qquad\text{in }\Hist(\Econ)
\]
is interpreted as a decomposed strategy, while
\[
  Q_{\Econ}(p):x\longrightarrow y
  \qquad\text{in }\Econ
\]
is its synthesized economic activity. If \(p,q:x\to y\) are histories, we
write
\[
  p\preceq q
\]
as an abbreviation for
\[
  Q_{\Econ}(p)\preceq Q_{\Econ}(q)
  \quad\text{in }\Econ(x,y).
\]

The natural transformation \(Q\) distinguishes two levels that will remain
important below.  A functor defined on \(\Econ\) can depend only on the
synthesized activity \(Q_{\Econ}(p)\), whereas a functor defined directly on
\(\Hist(\Econ)\) may retain the decomposition of the strategy.  The latter
possibility is needed for a path-dependent temporal realization.

\begin{remark}[Relation with the nerve]
A history of length \(n\) may equivalently be regarded as an \(n\)-simplex
of \(N_\bullet(\mathcal C)\). Thus the nerve decomposes histories according
to their lengths. We work primarily with \(\Hist(\mathcal C)\), because it
collects all finite histories in one category and has concatenation as its
composition. The nerve notation will not otherwise be used in this note.
\end{remark}

\section{Microscopic distortion on histories}
\label{sec:microscopic-distortion}

Let \(\Tmic\) be a microscopic time or context category and let
\[
  F:\Tmic^{\mathrm{op}}\longrightarrow\Prob
\]
be a categorical filtration. For \(m\in\Tmic\), write
\[
  F(m)=(\Omega_m,\mathcal F_m,\mu_m)
\]
and set
\[
  A_m:=L^0_{\geq0}(\Omega_m,\mathcal F_m,\mu_m).
\]
Thus \(A_m\) is the space of nonnegative random variables on \(F(m)\),
modulo almost-sure equality.

Put
\[
  \Hmic:=\Hist(\Tmic).
\]
We use \emph{Categorical Theory of Information Transport} (CTIT) for
the framework in which a categorical filtration supplies conditional
expectation transport along microscopic time arrows. Sections~3--5 of
\cite{AdachiHaraAB} develop the distortion, its integrability conditions,
and the resulting holonomy. The filtration-induced distortion is a typed
assignment on microscopic histories:
\[
  dF:
  \operatorname{Mor}(\Hmic)
  \longrightarrow
  \coprod_{m\in\Tmic}A_m.
  \tag{D}
\]
More precisely, if
\[
  \sigma:m\longrightarrow x
  \qquad\text{in }\Hmic,
\]
then
\[
  dF(\sigma)\in A_m.
  \tag{D1}
\]
The value is therefore represented on the probability space attached to the
initial microscopic object of the history.

The original construction of \(dF\) is degreewise on the nerve of \(\Tmic\)
\cite{AdachiHaraAB}.
A history of length \(n\) in \(\Hmic\) is the same data as an
\(n\)-simplex of \(N_\bullet(\Tmic)\), so equation \emph{(D)} simply collects
those degreewise values into one typed assignment on
\(\operatorname{Mor}(\Hmic)\). 

The assignment \(dF\) is not an ordinary functor to a fixed monoid: its
codomain varies with the initial object, and its multiplicative law involves
the transport supplied by \(F\). We use the established CTIT definition and
need only its typing, nonnegativity, and loop holonomy. In particular, we do
not assume that a nontrivial transport preserves the unit.

A microscopic AB-loop is a history
\[
  \gamma:m\longrightarrow m
  \qquad\text{in }\Hmic.
\]
Its holonomy distortion is
\[
  \Hol_F(\gamma):=dF(\gamma)\in A_m.
\]

\begin{definition}[Geometric AB-arbitrage]
The loop \(\gamma\) carries a geometric AB-arbitrage if
\[
  dF(\gamma)\geq\mathbf1_m
  \quad\mu_m\text{-almost surely},
  \qquad
  \mu_m\bigl(dF(\gamma)>\mathbf1_m\bigr)>0.
\]
\end{definition}

This is a geometric condition on a closed microscopic history. By itself, it
does not assert that an agent can choose, execute, finance, and settle a
corresponding economic strategy over a macro-time interval.

\section{The global unfolding as a twisted-arrow category}
\label{sec:globalUnfolding}

The history category \(\Hmic\) has a hom-functor
\[
  \operatorname{Hom}_{\Hmic}:
  \Hmic^{\mathrm{op}}\times\Hmic
  \longrightarrow\mathsf{Set}.
\]
We define the global unfolded microscopic category to be its category of
elements; this is the standard category-of-elements description of the
twisted-arrow category \cite[Tag 03JC]{Kerodon}:
\[
  \widetilde{\mathcal T}_{\mathrm{mic}}
  :=
  \Tw(\Hmic)
  \cong
  \int_{\Hmic^{\mathrm{op}}\times\Hmic}
  \operatorname{Hom}_{\Hmic}(-,-).
  \tag{TW}
\]

An object of \(\Tw(\Hmic)\) is a microscopic history
\[
  \alpha:m\longrightarrow x
  \qquad\text{in }\Hmic.
\]
A morphism
\[
  \xi=(u,v):\alpha\longrightarrow\beta
\]
from \(\alpha:m\to x\) to \(\beta:m'\to y\) consists of histories
\[
  u:m'\longrightarrow m,
  \qquad
  v:x\longrightarrow y
  \qquad\text{in }\Hmic
\]
satisfying
\[
  \beta=v\circ\alpha\circ u.
  \tag{F}
\]
Equivalently, in concatenation notation,
\[
  \beta=u*\alpha*v.
\]
Thus \(u\) changes the initial boundary contravariantly, while \(v\) adds a
terminal continuation covariantly.

There is a canonical endpoint projection
\[
  \lambda=(\src,\tgt):
  \Tw(\Hmic)
  \longrightarrow
  \Hmic^{\mathrm{op}}\times\Hmic.
\]
On objects,
\[
  \src(\alpha:m\to x)=m,
  \qquad
  \tgt(\alpha:m\to x)=x.
\]
On a morphism \(\xi=(u,v)\), the two components are
\[
  \src(\xi)=u^{\mathrm{op}},
  \qquad
  \tgt(\xi)=v.
  \tag{ET}
\]
The second equality will be essential when the distortion is recovered from
a twisted-arrow realization. For a fixed \(m\), the fiber of \(\src\) over
\(m\) is the coslice category
\[
  m\downarrow\Hmic.
\]

Let \(\gamma:m\to m\) be a loop in \(\Hmic\). Both
\[
  \varepsilon_m:m\to m
  \qquad\text{and}\qquad
  \gamma:m\to m
\]
are objects of \(\Tw(\Hmic)\). The factorization
\[
  \gamma
  =
  \gamma\circ\varepsilon_m\circ\varepsilon_m
\]
defines the canonical opening
\[
  \widehat\gamma
  =
  (\varepsilon_m,\gamma):
  \varepsilon_m\longrightarrow\gamma
  \qquad\text{in }\Tw(\Hmic).
  \tag{U}
\]
Its endpoints are distinct whenever \(\gamma\neq\varepsilon_m\), even though
both histories begin and end at \(m\). By \emph{(ET)}, its terminal component
is
\[
	\tgt(\widehat\gamma)=\gamma.
  \tag{UT}
\]

The target projection, followed by synthesis, gives the closing functor
\[
  C
  :=
  Q_{\Tmic}\circ\tgt:
  \Tw(\Hmic)\longrightarrow\Tmic.
\]
For the canonical opening,
\[
  C(\varepsilon_m)=C(\gamma)=m,
  \qquad
  C(\widehat\gamma)=Q_{\Tmic}(\gamma).
\]
Thus \(C\) closes the open arrow to the synthesized microscopic loop. The
unsynthesized terminal component \(\tgt(\widehat\gamma)=\gamma\), rather than
\(C(\widehat\gamma)\), is the history to which \(dF\) applies.

The category \(\Tw(\Hmic)\) therefore associates an open arrow canonically
with every based microscopic loop. The remaining question is whether an
observable macro-time history of an economic strategy can be realized by
such an opening.

\section{Observable macro time, microscopic realization, and distortion}
\label{sec:macro-micro-realization}

Recall the observable time projection \(\Pi:\Econ\to\Tmac\) introduced in
Section~\ref{sec:economy-category}.
Applying the history endofunctor gives
\[
  \Hist(\Pi):
  \Hist(\Econ)\longrightarrow\Hist(\Tmac),
  \tag{HP}
\]
which records the observable macro-time history of a decomposed economic
strategy.

The connection to microscopic geometry is supplied by a functor
\[
  J:
  \Hist(\Tmac)\longrightarrow\Tw(\Hmic).
  \tag{J}
\]
The functor \(J\) is a theoretical microscopic realization of macro-time
histories. It is part of the mathematical model and is not assumed to be
known or observable by an economic agent. Define
\[
  \widetilde\Pi
  :=J\circ\Hist(\Pi):
  \Hist(\Econ)\longrightarrow\Tw(\Hmic).
  \tag{E}
\]
Thus
\[
\begin{CD}
\Hist(\Econ) @>{\widetilde\Pi}>> \Tw(\Hmic)\\
@V{\Hist(\Pi)}VV @.\\
\Hist(\Tmac) @>>{J}> \Tw(\Hmic)
\end{CD}
\]
commutes by definition. The map \(\Pi\), rather than \(J\) or
\(\widetilde\Pi\), is the temporal information naturally available to market
participants.

The use of history categories remains substantive. We do not require
\(J\circ\Hist(\Pi)\) to factor through synthesis as
\[
  \widetilde\Pi=\Pi^\sharp\circ Q_{\Econ}
\]
for some \(\Pi^\sharp:\Econ\to\Tw(\Hmic)\). Such a factorization would make
the microscopic realization depend only on the synthesized economic
activity and would erase the decomposition of a strategy. The present
formulation allows it to depend on the observable macro-time path.

\subsection{The elementary one-loop realization problem}

Fix a geometric AB-loop \(\gamma:m\to m\) in \(\Hmic\) and its canonical
opening
\[
  \widehat\gamma:
  \varepsilon_m\longrightarrow\gamma
  \qquad\text{in }\Tw(\Hmic).
\]
We seek economic histories
\[
  p,q:x_0\longrightarrow x_1
  \qquad\text{in }\Hist(\Econ),
\]
where \(p\) is a benchmark and \(q\) is the candidate strategy. Their common
observable macro-time history is
\[
  \tau:t_0\longrightarrow t_1
  \qquad\text{in }\Hist(\Tmac),
\]
with
\[
  \Hist(\Pi)(p)=\Hist(\Pi)(q)=\tau.
  \tag{O}
\]
The macro history \(\tau\) realizes the opened microscopic loop when
\[
  J(t_0)=\varepsilon_m,
  \qquad
  J(t_1)=\gamma,
  \qquad
  J(\tau)=\widehat\gamma.
  \tag{R}
\]
Equations \emph{(O)}--\emph{(R)} imply
\[
  \widetilde\Pi(p)=\widetilde\Pi(q)=\widehat\gamma.
\]
Although \(\varepsilon_m\) and \(\gamma\) have the same microscopic boundary
object \(m\), they are distinct objects of \(\Tw(\Hmic)\) whenever
\(\gamma\neq\varepsilon_m\). The agent observes the macro interval
\(\tau:t_0\to t_1\), while \(J(\tau)=\widehat\gamma\) supplies its geometric
realization.

Condition \emph{(O)} may be weakened when only the synthesized macro interval
is observable. In that case one asks merely that
\[
  Q_{\Tmac}\bigl(\Hist(\Pi)(p)\bigr)
  =Q_{\Tmac}\bigl(\Hist(\Pi)(q)\bigr).
\]

The desired economic conclusion is
\[
  p\prec q,
\]
meaning
\[
  Q_{\Econ}(p)\prec Q_{\Econ}(q)
  \quad\text{in }\Econ(x_0,x_1).
\]
The benchmark \(p\) need not be unique. It is enough that at least one
benchmark makes \(q\) an economic fossil of the geometric AB-arbitrage.

\subsection{Distortion along the microscopic realization}

The morphism \(J(\tau)\) lies in \(\Tw(\Hmic)\), whereas \(dF\) is defined on
morphisms of \(\Hmic\). Therefore \(dF\) cannot be applied directly to
\(J(\tau)\). The target projection supplies the required microscopic
history:
\[
  J(\tau)\in\operatorname{Mor}(\Tw(\Hmic))
  \quad\xmapsto{\ \tgt\ }\quad
  \tgt(J(\tau))\in\operatorname{Mor}(\Hmic).
\]
We may consequently define the \(J\)-realized distortion of a macro history
\(\tau\) by
\[
  dF_J(\tau)
  :=
  dF\!\left(\tgt\bigl(J(\tau)\bigr)\right).
  \tag{DJ}
\]
This is a derived typed assignment on macro histories, not a redefinition of
\(dF\) and not a distortion whose domain is \(\Tw(\Hmic)\). If
\[
  \tgt(J(\tau)):m\longrightarrow x
  \qquad\text{in }\Hmic,
\]
then
\[
  dF_J(\tau)\in A_m.
\]

For the elementary one-loop realization, equations \emph{(R)} and
\emph{(UT)} give
\[
  \tgt(J(\tau))
  =
  \tgt(\widehat\gamma)
  =
  \gamma.
\]
Therefore
\[
	dF_J(\tau)=dF(\gamma).
  \tag{DL}
\]
The geometric AB-arbitrage condition may equivalently be written as
\[
  dF_J(\tau)\geq\mathbf1_m
  \quad\mu_m\text{-almost surely},
  \qquad
  \mu_m\bigl(dF_J(\tau)>\mathbf1_m\bigr)>0.
\]

The order of the constructions is important:
\[
  \Hist(\Tmac)
  \xrightarrow{\ J\ }
  \Tw(\Hmic)
  \xrightarrow{\ \tgt\ }
  \Hmic
  \xrightarrow{\ dF\ }
  \coprod_{m\in\Tmic}A_m.
\]
The closing functor \(C=Q_{\Tmic}\circ\tgt\) is used when a synthesized
microscopic arrow in \(\Tmic\) is needed. It is not used before evaluating
\(dF\), because synthesis could erase history-dependent information.

The realization datum may now be written
\[
  \mathfrak a=(F,\Pi,J,\gamma,p,q).
\]
Here \(\widehat\gamma\), \(\widetilde\Pi\), and \(dF_J\) are derived from the
displayed data. The observable clock \(\Pi\) belongs to the market
description, whereas \(J\) belongs to its geometric model and must be
specified independently of the favorable value of \(dF_J(\tau)=dF(\gamma)\).
Neither \(J\) nor \(\gamma\) is required to be recognized by the trader.

\section{Admissibility conditions and elementary tests}
\label{sec:admissibility}

\subsection{Provisional admissibility conditions}
\label{sec:admissibilityConditions}

The following conditions are candidates extracted from the elementary
example. They should initially be treated as a checklist to be tested, not as
a final irredundant axiom system.

\begin{definition}[Provisional admissible AB-realization]
The realization datum
\[
  \mathfrak a
  =
  (F,\Pi,J,\gamma,p,q)
\]
is admissible if it satisfies the following conditions.

\begin{enumerate}[label=\textnormal{(A\arabic*)},leftmargin=11mm]
\item \textbf{Observable temporal compatibility.}
The functor \(\Pi:\Econ\to\Tmac\) is part of the observable market
structure.  The benchmark and candidate determine the same macro-time
history \(\tau\):
\[
  \Hist(\Pi)(p)=\Hist(\Pi)(q)=\tau.
\]
When only the synthesized interval is observable, the weaker compatibility
condition following \emph{(R)} may be used instead.

\item \textbf{Geometric realizability.}
The loop is nontrivial, and the theoretical realization functor \(J\) is
specified independently of the favorable value of \(dF(\gamma)\).  It
satisfies
\[
  \gamma\neq\varepsilon_m,
  \qquad
  J(\tau)=\widehat\gamma:
  \varepsilon_m\longrightarrow\gamma.
\]
Neither \(J\) nor \(\gamma\) is assumed to be observable by the trader.

\item \textbf{Predictable selection.}
The decision rules governing the route, trading direction, and quantities
are specified before entry. Each action is selected using information
available immediately before that action; future realized quantities need
not be known at entry. For a fixed-history strategy, these rules implement
\(q\); branching policies are treated in Section~\ref{sec:executability}.  If
\(\mathcal G_{t_0-}\) denotes the trader's pre-entry information, the
entry decision must be \(\mathcal G_{t_0-}\)-measurable, and subsequent
decisions must satisfy the corresponding decision-time information
condition. The information available at successive decision times need not
be increasing; a policy may use earlier information only if it remains
available in the current subjective observation. The decision rule may
use \(\Pi\), but it need not use or recognize \(J\), \(\widetilde\Pi\), or
\(\gamma\).  The terminal gain need not be known at entry.

\item \textbf{Executable and complete economic realization.}
The histories \(p,q:x_0\to x_1\) have implementing policies executable
from specified initial configurations under the convention of
Section~\ref{sec:executability}. Almost-sure executability additionally
requires a specified stochastic realization.  Those interfaces and activities record every consequence
capable of affecting the comparison, including initial expenditure,
bid--ask spreads, transaction costs, financing, collateral, intermediate
cash flows, delivery, repayment, and terminal settlement.

\item \textbf{Benchmark compatibility and distortion realization.}
The histories \(p\) and \(q\) have the same economic interfaces, and the
realization identifies the geometric relative multiplier with the economic
advantage of \(q\) over \(p\). At minimum, the independently constructed
bridge should imply
\[
  dF(\gamma)\geq\mathbf1_m
  \quad\Longrightarrow\quad
  p\preceq q,
  \tag{B1}
\]
and, when the weak inequality holds,
\[
  \mu_m\bigl(dF(\gamma)>\mathbf1_m\bigr)>0
  \quad\Longrightarrow\quad
  p\prec q.
  \tag{B2}
\]
\end{enumerate}
\end{definition}

Condition (A5) is the least developed part of the definition. Merely assuming
\emph{(B1)}--\emph{(B2)} would make the desired conclusion formal. The real
mathematical problem is to construct the bridge from market data and then
prove that it has these order-preserving properties.

One possible stronger form is an external relative-return map
\[
  R_p:\mathcal A_p\longrightarrow A_m,
\]
defined on the class \(\mathcal A_p\) of strategies comparable with \(p\),
such that
\[
  R_p(p)=\mathbf1_m,
  \qquad
  R_p(q)=dF(\gamma),
\]
and such that \(R_p\) is order-reflecting on the realized strategies. The map
\(R_p\) is intended to translate an economic strategy into a random relative
performance multiplier, measured against the benchmark \(p\), and would be
additional realization data; it would not add algebraic operations to the
hom-sets of \(\Econ\).

\begin{proposition}[Formal consequence of admissibility]
Let \(\gamma:m\to m\) be a geometric AB-arbitrage. If \(\gamma\) admits
a provisional admissible realization relative to a benchmark \(p\), then
its realizing economic history \(q\) is a relative ordinal arbitrage over
\(p\):
\[
  p\prec q.
\]
\end{proposition}

\begin{proof}
The AB condition gives
\[
  dF(\gamma)\geq\mathbf1_m
  \quad\mu_m\text{-almost surely},
  \qquad
  \mu_m\bigl(dF(\gamma)>\mathbf1_m\bigr)>0.
\]
Condition (A5) transports weak and strict improvement to the economic
preorder, yielding \(p\prec q\).
\end{proof}

The proposition is deliberately elementary. It isolates the burden of the
theory: admissibility must be verified from causal, operational, and economic
data rather than postulated by restating \(p\prec q\).

\subsection{Small counterexamples as axiom tests}
\label{sec:axiomTests}

Each proposed condition rules out a different false positive.

\begin{center}
\begin{tabular}{p{0.61\textwidth} p{0.28\textwidth}}
\toprule
\textbf{Apparent opportunity} & \textbf{Failed condition}\\
\midrule
The benchmark and candidate do not determine the same observable macro-time
history. & Observable temporal compatibility\\
The observable macro history exists, but the proposed \(J\) does not realize
it as the canonical opening of the loop. & Geometric realizability\\
The profitable branch is selected only after future prices are observed.
& Predictable selection\\
Quoted exchange rates produce a favorable product, but the required trades
cannot be executed at those rates. & Executable realization\\
A positive receipt is accompanied by an omitted repayment or delivery
obligation. & Complete realization\\
The benchmark and candidate use different funding, settlement, or terminal
interfaces. & Benchmark compatibility\\
\bottomrule
\end{tabular}
\end{center}

These examples suggest that the admissibility problem has three layers:
\[
\boxed{
\begin{aligned}
&\text{observable--temporal:}&&
  \Pi,\ \tau,\ \mathcal G_{t_0-},\
  \text{ macro compatibility and predictability},\\
&\text{geometric:}&&
  \Hmic,\ \gamma,\ dF(\gamma),\ \widehat\gamma,\ J,\\
&\text{economic:}&&
  p,q,\preceq,\
  \text{ execution and complete liability-aware comparison}.
\end{aligned}}
\]

\section{Execution structures and executability}
\label{sec:executability}

Composability in $\Econ$ expresses compatibility of economic interfaces.
It does not imply that a proposed sequence can be executed with given
resources as information arrives. We now distinguish formal economic
histories, their realized executions, and policies specified before their
future outcomes are known.

\paragraph{The eight-step program.}
The following program organizes the construction. Section~\ref{sec:fxExample}
follows the same steps for a single foreign-exchange example.
\begin{enumerate}[label=\textnormal{\arabic*.},leftmargin=9mm]
\item Specify the economic interfaces, formal actions and observable clock.
\item Specify the execution projection, initial resources, capacities and
      possible outcomes.
\item Specify the readings available before each decision and the policies
      that act on those readings.
\item Realize the macro history microscopically and construct the
      probability spaces and compatible stochastic execution variables.
\item Use execution lifts and failure events to test executability.
\item Value fully settled terminal configurations and represent the
      policies economically.
\item Compute microscopic distortion, test the settlement bridge, and
      apply the compatibility with the economic preorder.
\item Test stability under identity insertion and harmless refinement.
\end{enumerate}

\subsection{The execution projection}
\label{subsec:execution-projection}

\begin{definition}[Execution structure]
An \emph{execution structure} over $\Econ$ is a category $\mathsf{Exec}$
equipped with a functor
\[
    p:\mathsf{Exec}\longrightarrow\Econ.
\]
Its objects are \emph{execution configurations}. A morphism
$\alpha:u\to v$ represents an execution of the economic action
$p(\alpha):p(u)\to p(v)$.
\end{definition}

A configuration may record holdings, remaining capacity, outstanding
commitments, and other circumstances relevant to execution. Its underlying
economic interface is $p(u)$ for a configuration $u$.

For an economic arrow $a:p(u)\to y$, write
\[
\operatorname{Lift}_u(a)
	:=
\{\alpha\in\operatorname{Mor}(\mathsf{Exec})
	\mid
\operatorname{dom}(\alpha)=u,\ p(\alpha)=a\}.
\]
We call $a$ \emph{locally feasible from $u$} if this set is nonempty.
Its members represent the possible executions of $a$, and their codomains
are the corresponding successor configurations. Existence and uniqueness
of lifts are not assumed. In particular, the mere existence of $a$ in
$\Econ$ does not guarantee its feasibility from a specified configuration.

If $u\xrightarrow{\alpha}v\xrightarrow{\beta}w$ are execution morphisms,
then
\[
    p(\beta\circ\alpha)=p(\beta)\circ p(\alpha).
\]
Thus executions compose when their full intermediate configuration matches.
This introduces no partial composition into either category. Instead,
resource compatibility is expressed by the existence of suitable execution
arrows. Capacities may depend on previous executions and need not be
encoded by the underlying economic interface alone.

\subsection{History lifting and realized execution}

The projection induces
\[
  \Hist(p):\Hist(\mathsf{Exec})\longrightarrow\Hist(\Econ).
\]
A formal history 
of $\Econ$
\[
\sigma=(
	x_0
		\xrightarrow{a_0}
	\cdots
		\xrightarrow{a_{n-1}}
	x_n
)
\]
has a \emph{realization from $u_0$}, where $p(u_0)=x_0$, if it admits a lift
\[
\eta = (
	u_0
		\xrightarrow{\alpha_0}
	\cdots
		\xrightarrow{\alpha_{n-1}}
	u_n
),
		\qquad
p(\alpha_k)=a_k.
\]
Histories retain intermediate configurations, whereas synthesis retains
only their composite. Naturality of $Q$ gives
\[
  p\circ Q_{\mathsf{Exec}}=Q_{\Econ}\circ\Hist(p).
\]
An execution of $Q_{\Econ}(\sigma)$ need not admit a factorization lifting
$\sigma$. Conversely, a lifted history always synthesizes to an execution
of its composite.

The existence of one lifted history establishes one possible realization.
It does not establish that a plan can be followed after every outcome of
its earlier actions. This distinction becomes essential when intermediate
holdings or capacities are random.

\subsection{Observations, policies, and pathwise executability}
\label{sec:ObservationsPolicies}

Fix an initial configuration $u_0$, a macro schedule
$\tau=(\delta_0,\ldots,\delta_{n-1})$, and prescribed economic interfaces
$x_0,\ldots,x_n$ over its successive dates. Let $\mathsf P_k$ be the set
of length-$k$ execution prefixes
\[
  \eta_k=(u_0\xrightarrow{\alpha_0}\cdots
                   \xrightarrow{\alpha_{k-1}}u_k)
\]
whose configurations project to the prescribed interfaces and whose
arrows satisfy $\Pi(p(\alpha_j))=\delta_j$ for $j<k$.
Here $\mathsf P_0$ consists of the empty prefix at $u_0$.
These sets are defined before choosing a policy.

\begin{definition}[Decision-time observation]
Assume that the execution prefixes in $\mathsf P_k$ have distinct
real labels, and identify $\mathsf P_k$ with the corresponding subset
of $\mathbb R$. For each $k<n$, let $J_k$ index the readings the trader
can consult immediately before decision $k$. Specify real-valued readouts
\[
  O_{k,j}:\mathsf P_k\longrightarrow\mathbb R,
  \qquad j\in J_k.
\]
The observation map and its value space are
\[
  \operatorname{obs}_k(\eta_k)
  :=\bigl(O_{k,j}(\eta_k)\bigr)_{j\in J_k},
  \qquad
  H_k:=\operatorname{obs}_k(\mathsf P_k)
       \subseteq\mathbb R^{J_k}.
\]
Equip $H_k$ with the trace $\mathcal H_k$ of the product of the Borel
$\sigma$-algebras on $\mathbb R^{J_k}$. For $J_k=\varnothing$, write $*$
for the empty tuple.
\end{definition}

The observation convention is part of the execution model. A reading
may be a displayed rate, a reported balance, or a record of an earlier
execution that the trader retains at decision $k$. A quote used at
decision $k$ must already be recorded in the prefix at that decision,
for example in the public-information component of $u_k$. The maps
$O_{k,j}$ specify which features of the prefix the trader can consult.
Two prefixes are observationally equivalent exactly when all of these
readings agree.

\begin{definition}[Policy with prescribed interfaces]
A policy of horizon $n$ is a family of maps
\[
  \pi_k:H_k\longrightarrow\Econ(x_k,x_{k+1}),
  \qquad k=0,\ldots,n-1.
\]
The action selected at $\eta_k$ is
$\pi_k(\operatorname{obs}_k(\eta_k))$, and clock compatibility requires
$\Pi(\pi_k(z))=\delta_k$ for every $z\in H_k$.
Prefixes with the same observation receive the same action.
The selected action need not be feasible from every configuration
compatible with that observation; feasibility is a separate
requirement.
\end{definition}

The policy is specified before entry; its selected quantities and actions may depend on later observations. 
A fixed formal history is the special case of
predetermined actions. In general a policy generates a branching family
of histories, not one fixed morphism of $\Hist(\Econ)$.

Fix $u_0$. The empty prefix at $u_0$ is reachable under $\pi$.
Recursively, a reachable prefix $\eta_k$ extends by any
\[
    \alpha_k\in\operatorname{Lift}_{u_k}(\pi_k(\operatorname{obs}_k(\eta_k))).
\]

\begin{definition}[Pathwise executability]
The policy $\pi$ is \emph{executable from $u_0$} if, for every reachable
prefix of length $k<n$,
\[
    \operatorname{Lift}_{u_k}(\pi_k(\operatorname{obs}_k(\eta_k)))\neq\varnothing.
\]
\end{definition}

This definition requires no probability measure. It demands the possibility
of continuation after every execution prefix generated by the policy.
For this interpretation, the outgoing lifts must include all outcomes
regarded as possible in the model, including adverse outcomes relevant to
subsequent feasibility. Nonempty lift sets alone would not describe
operational failure omitted from the model.

\subsection{Microscopic filtration and observable macro time}
\label{subsec:execution-filtration}

The distinction between the two time categories is essential. Recall
\[
    F:\Tmic^{\mathrm{op}}\longrightarrow\Prob,
    \qquad F(t)=(\Omega_t,\mathcal F_t,\mu_t),
\]
and the functors already introduced:
\[
    \Pi:\Econ\longrightarrow\Tmac,
    \qquad J:\Hist(\Tmac)\longrightarrow\Tw(\Hmic),
    \qquad \Hmic=\Hist(\Tmic).
\]

The observable clock of execution configurations is $\Pi\circ p$.
The microscopic realization of their histories is
\[
  J\circ\Hist(\Pi)\circ\Hist(p):
  \Hist(\mathsf{Exec})\longrightarrow\Tw(\Hmic).
\]
Let $C := Q_{\Tmic}\circ\tgt:\Tw(\Hmic)\to\Tmic$ be the closing functor
of Section~\ref{sec:globalUnfolding}.
Setting 
\[
K
	:=
C \circ J
	:
\Hist(\Tmac)
	\longrightarrow
\Tmic,
\] 
we obtain the induced filtration
\[
	F_{\mathrm{mac}}^J:=F\circ K^{\mathrm{op}}:
          \Hist(\Tmac)^{\mathrm{op}}\longrightarrow\Prob.
\]
Here ``macro'' refers to its domain of macro histories; descent to
$\Tmac$ itself would require an additional factorization through synthesis.

For a macro schedule
\[
    \tau=(\delta_0,\ldots,\delta_{n-1}),
    \qquad \delta_k:t_k\longrightarrow t_{k+1}
\]
in $\Hist(\Tmac)$,
write
\[
    m_k:=K(t_k),\qquad
    \kappa_k:=K((\delta_k)):m_k\longrightarrow m_{k+1}.
\]
The singleton parentheses distinguish an arrow of $\Tmac$ from its
one-step history. A policy following this schedule satisfies
$\Pi(a_k)=\delta_k$ at each stage.
The induced stagewise probability spaces are
\[
    F_{\mathrm{mac}}^J(t_k)=F(m_k)
       =(\Omega_{m_k},\mathcal F_{m_k},\mu_{m_k}).
\]
In particular, if $J(\tau)=\widehat\gamma$ with
$J(t_0)=\varepsilon_m$ and $J(t_n)=\gamma$, then
\[
    m_0=m_n=m,\qquad
    F_{\mathrm{mac}}^J(\tau)=F(Q_{\Tmic}(\gamma)).
\]
The endpoint probability spaces coincide, while the macro endpoints are
distinct and the transport need not be the identity.

\subsection{Observable information and stochastic execution data}

Equip the action hom-sets with measurable structures and require $\pi_k$
to be measurable. The stochastic realization supplies a measurable map
\[
  h_k:(\Omega_{m_k},\mathcal F_{m_k})\longrightarrow(H_k,\mathcal H_k).
\]
The information used by the policy on $\Omega_{m_k}$ is
$\mathcal G_k:=\sigma(h_k)=h_k^{-1}(\mathcal H_k)$.
A model may additionally specify a larger available information algebra
with respect to which $h_k$ is measurable.

For a finite-horizon realization, put
\[
   (\Omega_*,\mathcal F_*,\mu_*):=F(m_n).
\]
For $k\leq n$, let $\rho_k:\Omega_*\to\Omega_{m_k}$ be the measurable
map underlying the morphism supplied by $F$ along the remaining $K$-image
of the macro history. Thus $\rho_n=\id_{\Omega_*}$, and for $k<n$ it is
the underlying measurable map of
\[
F(\kappa_{n-1}\circ\kappa_{n-2}\circ\cdots\circ\kappa_k).
\]
Thus
\[
    \widehat h_k=h_k\circ\rho_k,
    \qquad \widehat a_k=\pi_k\circ\widehat h_k
\]
represent observations and decisions on one specified probability space.
Define the current subjective information on that space by
\[
  \mathcal I_k:=\rho_k^{-1}(\mathcal G_k)\subseteq\mathcal F_*.
\]
The family $(\mathcal I_k)$ need not be increasing, and
$\widehat a_k$ is $\mathcal I_k$-measurable. Thus the action is determined by the current observation. A decision cannot use
an earlier observation after it has disappeared from the current subjective
information.

If the maps in $\Prob$ are not measure-preserving, we do not identify
$(\rho_k)_*\mu_*$ with $\mu_{m_k}$; any transfer of almost-sure claims
between stages requires the appropriate null-set compatibility.

If one wishes to model a perfect-recall trader, one may instead use the
enlarged information
\[
  \mathcal I_k^{\mathrm{mem}}
  :=\sigma\!\left(\bigcup_{j=0}^k\mathcal I_j\right).
\]
A perfect-recall policy can use the tuple of observations up to decision
$k$ as its readout. If this tuple is to be represented on
$\Omega_{m_k}$, its common-space version must factor through $\rho_k$.

Further data are needed: measurable configuration and execution variables,
initial resources, and a rule for resolving the possible outgoing lifts.

On the common realization space $\Omega_*$, let
\[
  U_k:\Omega_*\to\operatorname{Ob}(\mathsf{Exec}),
  \qquad
  A_k:\Omega_*\to\operatorname{Mor}(\mathsf{Exec})
\]
be measurable random variables representing the configuration
before decision $k$ and the execution selected by the specified
outcome-resolution rule, respectively. The object and morphism
sets are equipped with specified measurable structures.
Before failure, they satisfy
\[
  \operatorname{dom}(A_k)=U_k,\qquad
  \operatorname{cod}(A_k)=U_{k+1},\qquad
  p(A_k)=\widehat a_k.
\]
On the survival events, the configuration $U_k$ is measurable with
respect to the trace of $\rho_k^{-1}(\mathcal F_{m_k})$. The completed
execution $A_k$ is measurable on the event of successful execution at
stage $k$ with respect to the trace of
$\rho_{k+1}^{-1}(\mathcal F_{m_{k+1}})$. This allows its outcome to be
recorded at the next stage. The choice of action uses the smaller
information $\mathcal I_k$ at the current stage.
For a fixed policy $\pi$, at a realization $\omega$ that reaches decision
$k$, let $\eta_k^\pi(\omega)$ be the prefix formed by the corresponding
realized executions. The observation compatibility condition is
\begin{equation}
  \operatorname{obs}_k\bigl(\eta_k^\pi(\omega)\bigr)
   =\widehat h_k(\omega)=h_k\bigl(\rho_k(\omega)\bigr).
  \label{eq:observation-compatibility}
\end{equation}
The prefix on the left is evaluated before execution of the selected
action at decision $k$. When comparing policies, the same $h_k$ may be
used if this condition holds for each policy; in general the realized
observations can depend on the policy through the execution prefixes.
Measurability and the choice of outcomes are additional stochastic
realization data; they are not consequences of the functor $p$ or of $F$.
A failure marker may be adjoined to keep the model defined after an
infeasible instruction.

\subsection{Almost-sure executability and concatenation}
\label{sec:AlmostSureExecutability}

Define the survival events and first-failure events recursively by
\[
  S_0:=\Omega_*,
\]
and, for $k=0,\ldots,n-1$,
\[
  B_k^\pi
  :=
  \left\{
    \omega\in S_k
\mid
    \operatorname{Lift}_{U_k(\omega)}
    \bigl(\pi_k(\widehat h_k(\omega))\bigr)
    =\varnothing
  \right\},
  \qquad
  S_{k+1}:=S_k\setminus B_k^\pi.
\]
The configuration $U_k$ is required on $S_k$.
On $S_{k+1}$, the specified outcome-resolution rule supplies
an execution lift and its successor configuration $U_{k+1}$.
The event of failure before the horizon is
\[
    B^\pi:=\bigcup_{k=0}^{n-1}B_k^\pi
          =\Omega_*\setminus S_n.
\]
We require these events to be measurable.
This requirement holds automatically in finite models equipped
with the full power-set sigma-algebras.


\begin{definition}[Almost-sure executability]
A policy is \emph{almost surely executable} in the specified stochastic
realization if
\[
    \mu_*(B^\pi)=0.
\]
\end{definition}

This is a property relative to the initial resources, observation rules,
and stochastic realization, not a property of a bare Econ history alone.
If a policy is pathwise executable, 
then every stochastic realization starting from the prescribed configuration 
and respecting the policy, observations, and execution lifts has an empty failure event. 
In particular, the policy is almost surely executable in every such realization.
The converse need not hold, since some possible prefixes may have probability zero.

Executable policies concatenate when the second is executable from every
configuration reachable at the end of the first, with compatible
observations. This follows by separating reachable prefixes into those
before and after the join. In the almost-sure version, one needs a common
stochastic realization and an almost-surely executable continuation at the
random joining configuration. Neither conclusion follows merely from
matching the underlying economic interfaces.

\section{Foreign-exchange policies and the settlement bridge}
\label{sec:fxExample}

We follow the eight steps of Section~\ref{sec:executability} for
a two-stage FX trade with a common initial stake of $w=100{,}000$ USD.
The benchmark converts through EUR and the candidate through GBP.
\subsection{Step 1. The dated FX diamond}

Consider two routes:
\[
  \mathrm{USD}\to\mathrm{EUR}\to\mathrm{JPY},
  \qquad
  \mathrm{USD}\to\mathrm{GBP}\to\mathrm{JPY}.
\]
Fix macro times
\[
  t_0<t_1<t_2
\]
and arrows
\[
  \delta_0:t_0\to t_1,
  \qquad
  \delta_1:t_1\to t_2.
\]
The first conversions are initiated at $t_0$, and their proceeds
are available before the second decision at $t_1$.
The second conversions settle by $t_2$.

Let $x_0,x_{\mathrm E},x_{\mathrm G},x_2$ be economic interfaces
with clock images
\[
  \Pi(x_0)=t_0,\qquad
  \Pi(x_{\mathrm E})=\Pi(x_{\mathrm G})=t_1,\qquad
  \Pi(x_2)=t_2.
\]
The intermediate interfaces permit EUR and GBP delivery,
respectively.  They specify boundary conditions, rather than
complete portfolios.

For an initial amount $w$ USD, consider formal economic actions
\[
  a_w:x_0\to x_{\mathrm E},
  \qquad
  c_w:x_0\to x_{\mathrm G},
\]
and
\[
  b_z:x_{\mathrm E}\to x_2,
  \qquad
  d_z:x_{\mathrm G}\to x_2.
\]
The first-leg rates are fixed:
\[
  a_w:\ w\ \mathrm{USD}\mapsto0.920w\ \mathrm{EUR},
\]
\[
  c_w:\ w\ \mathrm{USD}\mapsto0.790w\ \mathrm{GBP}.
\]
The actions $b_z$ and $d_z$ instruct conversion of $z$ units of
the intermediate currency at the prevailing terms of the specified
execution mechanism. Their execution arrows record the realized
rates and delivery consequences.

Assume
\[
  \Pi(a_w)=\Pi(c_w)=\delta_0,
  \qquad
  \Pi(b_z)=\Pi(d_z)=\delta_1.
\]
The two formal histories
\[
  p_w=(a_w,b_{0.920w}),
  \qquad
  q_w=(c_w,d_{0.790w})
\]
therefore satisfy
\[
  \Hist(\Pi)(p_w)
  =
  \Hist(\Pi)(q_w)
  =
  \tau:=(\delta_0,\delta_1).
\]

These are histories in $\Hist(\Econ)$.
Their feasibility from a particular configuration is determined
separately by execution lifts.

\subsection{Step 2. Initial resources, capacities, and outcomes}

Write $p_{\mathrm{ex}}:\mathsf{Exec}\to\Econ$ for the functor $p$
of Section~\ref{subsec:execution-projection} in this example. It sends each configuration to its
economic interface and each execution arrow to the economic action
it executes.

Fix an initial execution configuration $u_0$ with
\[
  p_{\mathrm{ex}}(u_0)=x_0.
\]
It records the initial $w$ USD, funding arrangements, and the
capacity commitments relevant to the example.

The stipulated input capacities are
\[
\begin{array}{c|r}
\text{Conversion}&\text{Input capacity}\\ \hline
\mathrm{USD}\to\mathrm{EUR}
  &1{,}000{,}000\ \mathrm{USD}\\
\mathrm{EUR}\to\mathrm{JPY}
  &900{,}000\ \mathrm{EUR}\\
\mathrm{USD}\to\mathrm{GBP}
  &800{,}000\ \mathrm{USD}\\
\mathrm{GBP}\to\mathrm{JPY}
  &600{,}000\ \mathrm{GBP}
\end{array}
\]
All required delivery and funding arrangements are included.
These capacities are available in both outcomes of the main example.

For an economic action $a:p_{\mathrm{ex}}(u)\to y$, recall
\[
  \operatorname{Lift}_u(a)
  =
  \left\{
    \alpha\in\operatorname{Mor}(\mathsf{Exec})
    \ \middle|\
    \operatorname{dom}(\alpha)=u,\ 
    p_{\mathrm{ex}}(\alpha)=a
  \right\}.
\]
An instruction exceeding the capacity recorded in $u$ has no
execution lift at the stated terms.

The EUR route requires
\[
  w\leq1{,}000{,}000,
  \qquad
  0.920w\leq900{,}000,
\]
and the GBP route requires
\[
  w\leq800{,}000,
  \qquad
  0.790w\leq600{,}000.
\]
Thus both routes are feasible under the stipulated capacities when
\[
  0<w\leq K_*,
\]
where
\[
  K_*=
  \min\left\{
    1{,}000{,}000,\,
    \frac{900{,}000}{0.920},\,
    800{,}000,\,
    \frac{600{,}000}{0.790}
  \right\}
  \approx759{,}493.67\ \mathrm{USD}.
\]

At $t_1$ the execution configuration records the intermediate holding
and the displayed pair of second-leg rates. The model supplies one
first-leg execution for each of the two possible displayed pairs, with
the same intermediate holding for a given route.

\subsection{Step 3. Observations and policies}

Before the first trade the empty prefix gives the constant observation $*$.
Before the second trade the readout is the displayed pair of conversion
rates, in JPY per EUR and JPY per GBP, respectively. Thus
\[
  H_0=\{*\},\qquad
  H_1=\{z_{\mathrm H},z_{\mathrm L}\},\qquad
  z_{\mathrm H}=(164.0,192.0),\quad z_{\mathrm L}=(160.0,187.0).
\]
The same readout applies on both routes.

The benchmark policy 
\[
\pi^p_0 : H_0 \to \Econ(x_0, x_{\mathrm E}),
	\quad
\pi^p_1 : H_1 \to \Econ(x_{\mathrm E}, x_2),
\]
is
\[
  \pi_0^p(*)=a_{100000},
  \qquad
  \pi_1^p(z_{\mathrm H})=\pi_1^p(z_{\mathrm L})=b_{92000}.
\]
The candidate policy
\[
\pi^q_0 : H_0 \to \Econ(x_0, x_{\mathrm G}),
	\quad
\pi^q_1 : H_1 \to \Econ(x_{\mathrm G}, x_2),
\]
is
\[
  \pi_0^q(*)=c_{100000},
  \qquad
  \pi_1^q(z_{\mathrm H})=\pi_1^q(z_{\mathrm L})=d_{79000}.
\]
These maps have the fixed hom-set codomains prescribed by their
respective interface sequences, and
\[
  \Pi(\pi_k^r(h))=\delta_k,
  \qquad r\in\{p,q\}.
\]

The routes are selected before entry.
The second-stage instruction is the same in both outcomes:
convert the specified intermediate holding at the prevailing
terms of the execution mechanism.
The execution arrow, rather than the policy action alone,
records the realized rate.

\subsection{Step 4. Microscopic realization and stochastic execution}

\paragraph{Macro schedule and microscopic loop.}
Let
\[
  \Tmac=[2]
  =\{t_0\xrightarrow{\delta_0}t_1
                 \xrightarrow{\delta_1}t_2\},
\]
and write
\[
  \delta_{02}:=\delta_1\circ\delta_0.
\]

Let $\Tmic$ be the free category generated by the three-leg cycle
\[
  m_0\xrightarrow{i_1}m_1
  \xrightarrow{i_2}m_2
  \xrightarrow{i_3}m_0.
\]
In $\Hmic=\Hist(\Tmic)$, put
\[
  \eta=(i_1):m_0\to m_1,
  \qquad
  \kappa=(i_2,i_3):m_1\to m_0,
\]
and
\[
  \gamma:=\kappa\circ\eta
  =(i_1,i_2,i_3):m_0\to m_0.
\]
This is the microscopic loop used in the parameterized version of
Example~7.2 of \cite{AdachiHaraAB} below.

Define
\[
  J_0:\Tmac\longrightarrow\Tw(\Hmic)
\]
on objects by
\[
  J_0(t_0)=\varepsilon_{m_0},
  \qquad
  J_0(t_1)=\eta,
  \qquad
  J_0(t_2)=\gamma.
\]
On the consecutive arrows, define
\[
  J_0(\delta_0)
  =
  (\varepsilon_{m_0},\eta):
  \varepsilon_{m_0}\to\eta,
\]
\[
  J_0(\delta_1)
  =
  (\varepsilon_{m_0},\kappa):
  \eta\to\gamma.
\]
Indeed, $\gamma=\kappa\circ\eta\circ\varepsilon_{m_0}$, so the
second displayed pair is a well-typed twisted-arrow morphism. Their
composite is
\[
  J_0(\delta_{02})
  =
  (\varepsilon_{m_0},\gamma)
  =
  \widehat\gamma.
\]

The functor required in Section~\ref{sec:executability} is
\[
  J:=J_0\circ Q_{\Tmac}:
  \Hist(\Tmac)\longrightarrow\Tw(\Hmic).
\]
Let
\[
  \tau := (\delta_0,\delta_1)
  \in \Hist(\Tmac)(t_0,t_2)
\]
denote the two-step macro-time history. Then
\[
  Q_{\Tmac}(\tau)=\delta_{02},
  \qquad
  J(\tau)=\widehat\gamma.
\]
Thus the first macro interval realizes the microscopic leg $i_1$, while
the second macro interval realizes the two-leg microscopic continuation
$(i_2,i_3)$. The macro schedule need not have the same number of stages
as its microscopic realization.

Using
\[
  C=Q_{\Tmic}\circ\tgt:
  \Tw(\Hmic)\to\Tmic,
  \qquad
  K=C\circ J:
  \Hist(\Tmac)\to\Tmic,
\]
we obtain
\[
  K(t_0)=m_0,
  \qquad K(t_1)=m_1,
  \qquad K(t_2)=m_0,
\]
and
\[
  K((\delta_0))=i_1,
  \qquad
  K((\delta_1))=i_3\circ i_2.
\]

\paragraph{Filtration and observation maps.}
\label{sec:parameterized-non-flat-filtration}

We now define a microscopic filtration
$
  F:\Tmic^{\mathrm{op}}\to\Prob
$
by parameterizing Example~7.2 of \cite{AdachiHaraAB}.
Let
\[
  \lambda_U>0,
  \qquad
  \lambda_D>0,
  \qquad
  \frac{\lambda_U}{2}+\frac{\lambda_D}{4}<1.
  \tag{P1}
\]
The parameters will later be matched with the two statewise relative
settlement multipliers of the FX policies.

At the three microscopic objects, set
\[
  F(m_0)
  =
  \bigl(\{U,D\},2^{\{U,D\}},\mu_0\bigr),
  \qquad
  \mu_0(U)=\mu_0(D)=\frac12,
\]
\[
  F(m_1)
  =
  \bigl(\{a,b,c\},2^{\{a,b,c\}},\mu_1\bigr),
\]
where
\[
  \mu_1(a)=\frac12,
  \qquad
  \mu_1(b)=\mu_1(c)=\frac14,
\]
and
\[
  F(m_2)
  =
  \bigl(\{u,v,w\},2^{\{u,v,w\}},\mu_2\bigr),
\]
where
\[
  \mu_2(u)=\frac{\lambda_U}{2},
  \qquad
  \mu_2(w)=\frac{\lambda_D}{4},
  \qquad
  \mu_2(v)=1-\frac{\lambda_U}{2}-\frac{\lambda_D}{4}.
  \tag{P2}
\]
Condition \emph{(P1)} makes all three masses in \emph{(P2)} strictly
positive.

The measurable maps underlying the contravariant functor
\[
  F:\Tmic^{\mathrm{op}}\longrightarrow\Prob
\]
are
\[
\begin{array}{rclcrcl}
F(i_1):\{a,b,c\}&\longrightarrow&\{U,D\},
&&a&\longmapsto&U,\\
&&&&b,c&\longmapsto&D,
\end{array}
\]
\[
\begin{array}{rclcrcl}
F(i_2):\{u,v,w\}&\longrightarrow&\{a,b,c\},
&&u&\longmapsto&a,\\
&&&&v&\longmapsto&c,\\
&&&&w&\longmapsto&b,
\end{array}
\]
and
\[
  F(i_3):\{U,D\}\longrightarrow\{u,v,w\},
  \qquad
  U\longmapsto u,
  \quad
  D\longmapsto w.
\]
These assignments determine $F$ on the free category $\Tmic$.

The common realization space of Section~\ref{sec:executability} is
\[
  (\Omega_*,\mathcal F_*,\mu_*)
  =F(m_0)
  =\bigl(\{U,D\},2^{\{U,D\}},\mu_0\bigr).
\]
The remaining-history maps are
\[
  \rho_0=\id_{\{U,D\}},
  \qquad
  \rho_2=\id_{\{U,D\}},
\]
and
\[
  \rho_1:\{U,D\}\longrightarrow\{a,b,c\},
  \qquad
  \rho_1(U)=a,
  \quad
  \rho_1(D)=b.
  \tag{O1}
\]
Indeed, $\rho_1$ is the measurable map underlying
$F(i_3\circ i_2)$, and the full-loop map underlying
$F(i_3\circ i_2\circ i_1)$ is the identity on $\{U,D\}$.

The observation maps into the value spaces from Step~3 are
\[
  h_0(U)=h_0(D)=*,\qquad
  h_1(a)=z_{\mathrm H},\qquad h_1(b)=h_1(c)=z_{\mathrm L}.
\]
Consequently $\mathcal G_0$ is trivial and $\mathcal G_1$ is generated by
the partition $\{\{a\},\{b,c\}\}$. On the common realization space,
$\widehat h_0$ is constant and
\[
  \widehat h_1(U)=z_{\mathrm H},
  \qquad \widehat h_1(D)=z_{\mathrm L}.
\]
The displayed quotes distinguish the two realized outcomes on $\Omega_*$.
At the microscopic stage $m_1$, the same quote pair is displayed at $b$
and $c$.

\paragraph{Stochastic execution variables.}
For each route $r\in\{p,q\}$ and state $\omega\in\{U,D\}$, specify
configurations and execution arrows
\[
  u_0\xrightarrow{\alpha_0^{r,\omega}}u_1^{r,\omega}
       \xrightarrow{\alpha_1^{r,\omega}}u_2^{r,\omega}.
\]
The two first-stage arrows for a fixed route have the same economic
projection and the same currency delivery, while their codomains record the
pair of quotes displayed at $t_1$.
Their projections are
\[
  p_{\mathrm{ex}}(\alpha_0^{p,\omega})=a_{100000},
  \qquad
  p_{\mathrm{ex}}(\alpha_0^{q,\omega})=c_{100000},
\]
\[
  p_{\mathrm{ex}}(\alpha_1^{p,\omega})=b_{92000},
  \qquad
  p_{\mathrm{ex}}(\alpha_1^{q,\omega})=d_{79000}.
\]
The configurations $u_1^{p,\omega},u_1^{q,\omega}$ record holdings of
$92{,}000$ EUR and $79{,}000$ GBP, respectively, together with the displayed
pair of quotes for $\omega$.
The second execution arrows record the JPY deliveries at those rates.

These specifications define the random variables of 
Section~\ref{sec:executability};
\[
  U_0^r(\omega)=u_0,\qquad
  A_0^r(\omega)=\alpha_0^{r,\omega},\qquad
  U_1^r(\omega)=u_1^{r,\omega},
\]
\[
  A_1^r(\omega)=\alpha_1^{r,\omega},
  \qquad
  U_2^r(\omega)=u_2^{r,\omega}.
\]
They satisfy
\[
  \operatorname{dom}(A_k^r)=U_k^r,\qquad
  \operatorname{cod}(A_k^r)=U_{k+1}^r,
\]
and
\[
  p_{\mathrm{ex}}(A_k^r(\omega))
  =
  \pi_k^r(\widehat h_k(\omega)).
\]
Equip the finite configuration and execution collections with
their power-set sigma-algebras.
All these maps are then measurable.

The execution prefix at decision $1$ displays $z_{\mathrm H}$ for
$\omega=U$ and $z_{\mathrm L}$ for $\omega=D$. Its readout agrees with
$h_1\circ\rho_1$ on both routes. At decision $0$ both readouts are constant,
so \eqref{eq:observation-compatibility} holds at both decisions.
For the stagewise measurability requirements, $U_0^r$ is constant,
while $A_0^r$ and $U_1^r$ factor through $\rho_1$. The factor maps can
use the low-quote execution and configuration at both $b$ and $c$.
The second execution $A_1^r$ and its terminal configuration $U_2^r$
are measurable at the terminal stage, where $\rho_2$ is the identity.

\subsection{Step 5. Lift sets and failure events}

The execution structure includes both second-stage outcomes and
no additional adverse outcomes in this illustrative model.
All selected quantities fit within capacity in both states.
Therefore each policy has a continuation from every reachable
prefix and is pathwise executable.

In the notation of 
Section \ref{sec:AlmostSureExecutability},
\[
  B_0^{\pi^r}=B_1^{\pi^r}=\varnothing,
  \qquad
  B^{\pi^r}=\varnothing,
  \qquad r\in\{p,q\}.
\]
In particular, both policies are almost surely executable.

\paragraph{Capacity-failure variant.}
Keep the two equally likely outcomes and suppose the GBP holding
$Y=\operatorname{gbp}\circ U_1$ after the first trade is
$Y(U)=79{,}000$ and $Y(D)=81{,}000$ GBP. The second trade has an input
capacity of $80{,}000$ GBP. Use the displayed GBP balance as the readout
in this variant, with $H_1^{\mathrm{bal}}=\{79{,}000,81{,}000\}$.
On reached prefixes
\[
  \operatorname{obs}^{\mathrm{bal}}_1(\eta_1^\pi(\omega))
  =\widehat h^{\mathrm{bal}}_1(\omega)=Y(\omega).
\]
For example the corresponding microscopic readout on $a,b,c$ takes
values $79{,}000,81{,}000,81{,}000$, respectively, and
$\widehat h^{\mathrm{bal}}_1=h^{\mathrm{bal}}_1\circ\rho_1$.
The policy $\pi_1(z)=d_z$ specifies full conversion before entry, with
the amount selected when the balance is read. If the first trade is
feasible in both outcomes, then
\[
  \operatorname{Lift}_{U_1(U)}(d_{79000})\ne\varnothing,\qquad
  \operatorname{Lift}_{U_1(D)}(d_{81000})=\varnothing.
\]
Thus $B^\pi=\{D\}$ with $\mu_*(B^\pi)=1/2$. A policy that caps conversion
at $80{,}000$ GBP can continue whenever the capped lifts exist.
Its settlement valuation must account for the residual GBP.

\subsection{Step 6. Settlement and economic representation}

Let
\[
  \operatorname{Term}(x_2)
  :=
  \left\{
    u\in\operatorname{Ob}(\mathsf{Exec})
    \ \middle|\
    p_{\mathrm{ex}}(u)=x_2
    \text{ and }u\text{ is fully settled}
  \right\}.
\]
Define the settlement valuation
\[
  v_{\mathrm{JPY}}:
  \operatorname{Term}(x_2)\longrightarrow\mathbb R
\]
to be the net JPY value of a fully settled terminal configuration,
including all relevant assets, costs, payments, and liabilities.

For each route $r\in\{p,q\}$, the random
terminal configuration is
\[
  U_2^r:\Omega_*\longrightarrow\operatorname{Term}(x_2).
\]
Define its random settlement value by
\[
  V_*(\pi^r)
  :=
  v_{\mathrm{JPY}}\circ U_2^r:
  \Omega_*\longrightarrow\mathbb R.
\]
Equivalently,
\[
  V_*(\pi^r)(\omega)
  =
  v_{\mathrm{JPY}}\bigl(U_2^r(\omega)\bigr).
\]
This is the explicit connection between the random execution
variables of 
Section~\ref{sec:executability}
and the settlement valuation.

With no additional costs,
\[
\begin{array}{c|rrr}
 &V_*(\pi^p)&V_*(\pi^q)
 &V_*(\pi^q)-V_*(\pi^p)\\ \hline
 U&15{,}088{,}000&15{,}168{,}000&80{,}000\\
 D&14{,}720{,}000&14{,}773{,}000&53{,}000
\end{array}
\]
All entries are in JPY.
The values belong to
\[
  \mathcal V_*
  :=
  L^0(\Omega_*,\mathcal F_*,\mu_*;\mathbb R).
\]
The general codomain is signed, although the values in this
example are positive.

The relative multiplier is
\[
  R_{\pi^p}(\pi^q)
  :=
  \frac{V_*(\pi^q)}{V_*(\pi^p)}
  =
  \begin{cases}
    15168/15088,&U,\\
    14773/14720,&D.
  \end{cases}
\]
It exceeds $1$ in both states.

The policies in this example select fixed economic actions,
even though their executions have random consequences.
Thus their projected histories are the fixed histories
$p_w,q_w$.
We may represent the policies economically by
\[
  e(\pi^p)=Q_{\Econ}(p_w),
  \qquad
  e(\pi^q)=Q_{\Econ}(q_w).
\]
If the hom-preorder recognizes statewise larger net JPY delivery
as strictly preferable when other relevant consequences agree,
then
\[
  e(\pi^p)\prec e(\pi^q).
\]

This conclusion uses compatibility of the external valuation
with the economic preorder.
It is not a consequence of the execution projection alone.
For policies whose selected economic actions genuinely branch,
a representation $e(\pi)$ requires additional construction;
synthesis applies directly only to individual histories.

\subsection{Step 7. Distortion and the realized settlement bridge}

We record the distortion calculation because it is the essential feature
of the construction. The one-step distortions are
\[
  dF(i_1)=\mathbf 1_{m_0},
\]
\[
  dF(i_2)(a)=\lambda_U,
  \qquad
  dF(i_2)(b)=\lambda_D,
  \qquad
  dF(i_2)(c)=4-2\lambda_U-\lambda_D,
\]
and
\[
  dF(i_3)(u)=\frac1{\lambda_U},
  \qquad
  dF(i_3)(v)=0,
  \qquad
  dF(i_3)(w)=\frac2{\lambda_D}.
\]
The recursive rule gives tail distortion
$(\lambda_U,2\lambda_D,0)$ on $(a,b,c)$ for the history $(i_2,i_3)$.
The first transport takes the value at $a$ for $U$, and averages the
values at $b,c$ with equal weights for $D$. Since $dF(i_1)=\mathbf 1_{m_0}$,
the loop $\gamma=(i_1,i_2,i_3)$ has distortion
\[
  dF(\gamma)(U)=\lambda_U,
  \qquad
  dF(\gamma)(D)=\lambda_D.
  \tag{P3}
\]
Thus
\[
  dF(\gamma)
  =
  \lambda_U\mathbf1_{\{U\}}
  +\lambda_D\mathbf1_{\{D\}}
  \in A_{m_0}.
  \tag{P4}
\]
Notice that the local factors do not simply multiply pointwise: each
factor is transported to the preceding probability space before the next
multiplication. This is precisely the mechanism exhibited by
Example~7.2 of \cite{AdachiHaraAB}.

We now specialize the parameters of the filtration to the relative
settlement multipliers calculated above:
\[
  \lambda_U
  :=\frac{V_*(\pi^q)(U)}{V_*(\pi^p)(U)}
  =\frac{15168}{15088}
  =\frac{948}{943},
\]
\[
  \lambda_D
  :=\frac{V_*(\pi^q)(D)}{V_*(\pi^p)(D)}
  =\frac{14773}{14720}.
  \tag{FX1}
\]
Both parameters are strictly greater than $1$, and they satisfy
condition \emph{(P1)}. In particular, the intermediate measure in
\emph{(P2)} is
\[
  \mu_2(u)=\frac{474}{943},
  \qquad
  \mu_2(w)=\frac{14773}{58880},
  \qquad
  \mu_2(v)=\frac{594947}{2414080}.
\]
These three numbers are positive and sum to $1$.

The calculation in \emph{(P3)}--\emph{(P4)} uses the microscopic filtration. After the specialization \emph{(FX1)}, it agrees
statewise with the independently evaluated relative settlement multiplier:
\[
  dF(\gamma)
  =R_{\pi^p}(\pi^q)
  =\frac{V_*(\pi^q)}{V_*(\pi^p)}.
  \tag{FX2}
\]
Because $V_*(\pi^p)>0$ in both states, equation \emph{(FX2)} is equivalent
to the settlement bridge
\[
  V_*(\pi^q)=dF(\gamma)V_*(\pi^p)
  \qquad \mu_*\text{-almost surely}.
  \tag{FX3}
\]
Moreover,
\[
  dF(\gamma)>\mathbf1_{m_0}
  \qquad \mu_*\text{-almost surely},
\]
so $\gamma$ is a geometric AB-arbitrage. The order compatibility assumed
above therefore gives
\[
  e(\pi^p)\prec e(\pi^q).
\]
The two-stage FX relative arbitrage is consequently represented as an
economic fossil of the three-leg microscopic AB-loop.

This construction resolves the finite representability problem: the
parameterized family contains a filtration whose distortion realizes the
observed two-state economic multiplier. It does not yet explain why the
microscopic masses in \emph{(P2)} should be selected by market primitives
independently of the settlement values. 
Thus the remaining problem is to establish the naturality and economic
meaning of the coupling that determines these parameters.

\begin{proposition}[Settlement bridge criterion]
Fix an initial configuration \(u_0\) over \(x_0\), a macro
schedule \(\tau\), a terminal interface \(x_2\), and compatible
observation and stochastic execution data. Let
\[
  \mathcal P=\mathcal P(u_0,\tau,x_2)
\]
denote the set of policies that begin from $u_0$, follow the macro schedule
$\tau$, are almost surely executable in the fixed stochastic realization,
and have fully settled terminal configurations over $x_2$. Suppose
$\pi^p,\pi^q\in\mathcal P$.

For each \(\pi\in\mathcal P\), let
\[
  U_2^\pi:\Omega_*\longrightarrow\operatorname{Term}(x_2)
\]
be its random terminal configuration. Define
\[
  V_*(\pi)
  :=
  v_{\mathrm{JPY}}\circ U_2^\pi
  \in
  \mathcal V_*,
\]
where
\[
  \mathcal V_*
  :=
  L^0(\Omega_*,\mathcal F_*,\mu_*;\mathbb R).
\]
This gives a valuation map
\[
  V_*:\mathcal P\longrightarrow\mathcal V_*.
\]
Suppose that the policies under consideration admit an
economic representation
\[
  e:\mathcal P\longrightarrow\Econ(x_0,x_2).
\]
Suppose also that the macro history is realized by
$J(\tau)=\widehat\gamma$, and assume
\[
  V_*(\pi^p)>0\quad\mu_*\text{-almost surely},
\]
\[
  V_*(\pi^q)=dF(\gamma)V_*(\pi^p).
\]
On the represented policy class, assume
\[
  V_*(\pi)\leq V_*(\pi')
  \quad\mu_*\text{-almost surely}
  \quad\Longrightarrow\quad
  e(\pi)\preceq e(\pi'),
\]
and that this weak inequality together with
\[
  \mu_*\bigl(V_*(\pi)<V_*(\pi')\bigr)>0
\]
implies $e(\pi)\prec e(\pi')$.
If $\gamma$ is a geometric AB-arbitrage, then
\[
  e(\pi^p)\prec e(\pi^q).
\]
\end{proposition}

\begin{proof}
The bridge identity and positivity of the benchmark give
\[
  V_*(\pi^q)\geq V_*(\pi^p)
  \quad\mu_*\text{-almost surely}.
\]
The inequality is strict on the positive-measure event
$\{dF(\gamma)>\mathbf1_m\}$.
The assumed compatibility with the hom-preorder gives the result.
\end{proof}

The benchmark need not be unique.
For strictly positive benchmark values under the same valuation
conventions,
\[
  R_{\pi^{p'}}(\pi)
  =
  \frac{R_{\pi^p}(\pi)}
       {R_{\pi^p}(\pi^{p'})}.
\]
All arithmetic here belongs to external settlement valuation;
no additive structure is imposed on Econ hom-sets.

\subsection{Step 8. Stability under refinement}

A refinement test compares the execution observations, settlement values,
and microscopic distortion before and after inserting an identity or
subdividing a stage. The relevant refinements are those that preserve
the economic schedule at its original decision dates and supply the
same observations and execution opportunities there. Whether the bridge
is stable under such refinements remains open. A separate open problem
from Step~7 is to determine the filtration parameters from market
primitives independently of the settlement ratios.

\section{Conclusion and outlook}
\label{conclusion}

This note has addressed the following question: when can a geometric
AB-arbitrage, expressed by the distortion of a closed microscopic
history, be realized as an executable relative arbitrage over a genuine
macro-time interval?  The question contains three logically distinct
parts.  The microscopic filtration must produce a favorable distortion,
the corresponding economic strategy must be executable using information
available at the time of each decision, and the microscopic distortion
must agree with the relative economic performance of the realized
strategy.  Keeping these parts separate is necessary because none of
them follows from the other two.

The categorical framework developed here records this separation.  The
preorder-enriched category \(\Econ\) describes economic activities under
common boundary interfaces, while \(\Hist(\Econ)\) retains their staged
decompositions.  The observable clock
\[
  \Pi:\Econ\longrightarrow\Tmac
\]
extends to histories and records the macro-time schedule visible to a
market participant.  On the microscopic side, the filtration
\(F:\Tmic^{\mathrm{op}}\to\Prob\) assigns a typed distortion
\(dF(\sigma)\) to a microscopic history \(\sigma\).  The twisted-arrow
category \(\Tw(\Hmic)\) makes it possible to represent a closed
microscopic loop by an open arrow between histories.  A realization
functor
\[
  J:\Hist(\Tmac)\longrightarrow\Tw(\Hmic)
\]
then relates an observable macro history to its microscopic geometry.
The resulting distortion along a macro history is the derived quantity
\[
  dF_J(\tau)
  :=dF\bigl(\tgt(J(\tau))\bigr).
\]
In the elementary one-loop case, this construction gives
\(dF_J(\tau)=dF(\gamma)\).  Thus the use of a twisted-arrow category does
not change the domain of \(dF\); it supplies the microscopic history on
which \(dF\) is evaluated.

The execution structure gives a mathematical interpretation of the
execution requirement in admissibility.  Formal economic histories are
lifted to histories of configurations subject to resource and capacity
constraints.  In a stochastic setting, observations and policies specify
which actions may be chosen from the information available at each macro
time.  This leads to distinct notions of pathwise and almost-sure
executability.  It also shows why executability cannot in general be
identified with the mere existence of an arrow in \(\Econ\): a formal
activity may fail to admit a feasible lift for some configurations or on
a positive-probability event.  Moreover, closure under concatenation and
closure under taking subhistories are additional properties to be checked,
rather than automatic consequences of the definitions.

The finite foreign-exchange example illustrates the first seven steps of
the proposed program.  It displays explicitly the macro schedule,
capacity-constrained configurations, observations, policies, stochastic
execution lifts, terminal settlement, and the economic representation of
the resulting policies.  It also clarifies the role of a benchmark.  
A candidate policy is compared with any suitable benchmark policy having
the same economic interfaces, and the benchmark need not be unique.

The parameterized non-flat filtration advances the settlement problem.
Its holonomy is
\[
  dF(\gamma)
  =\lambda_U\mathbf1_{\{U\}}
   +\lambda_D\mathbf1_{\{D\}},
\]
and choosing the two parameters to equal the statewise FX settlement ratios
gives
\[
  V_*(\pi^q)=dF(\gamma)V_*(\pi^p).
\]
Together with compatibility between settlement-value dominance and the
hom-preorder of \(\Econ\), this implies
\[
  e(\pi^p)\prec e(\pi^q).
\]
Thus the finite model contains an executable economic fossil of a geometric
AB-arbitrage and completes Step~7 as a representability result.

The construction also sharpens what remains unresolved.  The microscopic
masses were calibrated so that their holonomy agrees with the settlement
ratio.  A stronger theory should derive or identify those masses from
market primitives, execution data, or another independently motivated
coupling.  It should also determine when the bridge survives identity
insertion and harmless refinement of histories.  These naturality and
stability questions, rather than the bare existence of a finite bridge,
are the next problems for the admissibility theory.

\end{document}